\RequirePackage{fix-cm}
\documentclass[natbib,smallextended]{svjour3}       % onecolumn (second format)
\smartqed  % flush right qed marks, e.g. at end of proof
\usepackage{graphicx}
 \journalname{my journal}
\def\t0{\theta_{\circ}}

\def\be{\begin{equation}}
\def\en{\end{equation}}

\def\msun{M_{\odot}}

\def\lsun{L_{\odot}}
\def\mdot{\dot{M}}
\def\msunyr{M_{\odot}\, {\rm yr^{-1}}}

\def\etal{{\it et al.\ }}
\def\gcm2{\, \rm g\, cm^{-2}}
\def\nus0{\nu_{\circ}}

\def\R0{R_{\circ}}

\newbox\grsign \setbox\grsign=\hbox{$>$} \newdimen\grdimen \grdimen=\ht\grsign
\newbox\simlessbox \newbox\simgreatbox
\setbox\simgreatbox=\hbox{\raise.5ex\hbox{$>$}\llap
     {\lower.5ex\hbox{$\sim$}}}\ht1=\grdimen\dp1=0pt
\setbox\simlessbox=\hbox{\raise.5ex\hbox{$<$}\llap
     {\lower.5ex\hbox{$\sim$}}}\ht2=\grdimen\dp2=0pt
\def\simgreat{\mathrel{\copy\simgreatbox}}
\def\simless{\mathrel{\copy\simlessbox}}

\usepackage{color,ulem,soul}

\definecolor{dark-red}{rgb}{0.75, 0.00, 0.00}
\definecolor{hlcolor}{rgb}{1.00, 0.90, 0.85}
\sethlcolor{hlcolor}

\renewcommand\emph[1]{\textit{#1}}
\makeatletter
\DeclareRobustCommand{\em}{%
  \@nomath\em \if b\expandafter\@car\f@series\@nil
  \normalfont \else \itshape \fi}
\makeatother

\begin{document}

\title{Disk evolution and the fate of water %\thanks{Grants or other notes
%about the article that should go on the front page should be
%placed here. General acknowledgments should be placed at the end of the article.}
}
%\subtitle{Do you have a subtitle?\\ If so, write it here}

%\titlerunning{Short form of title}        % if too long for running head

\author{Lee~Hartmann         \and
        Fred~Ciesla          \and
        Oliver~Gressel    \and
        Richard~Alexander    %etc.
}

%\authorrunning{Short form of author list} % if too long for running head

\institute{ L.~Hartmann \at
              Dept. of Astronomy, University of Michigan, Ann Arbor, MI USA\\
              %Tel.: +1-936-7781\\
              %Fax: +1-763-6317\\
              \email{lhartm@umich.edu}           %  \\
%             \emph{Present address:} of F. Author  %  if needed
            \and
            F.~Ciesla \at
              Dept. Geophysical Sciences, University of Chicago, Chicago, IL USA\\
              \email{fciesla@uchicago.edu}
            \and
            O.~Gressel \at
            The Niels Bohr Institute, Blegdamsvej 17, 2100 Copenhagen \O, Denmark\\
            \email{gressel@nbi.ku.dk}
            \and
            R.~Alexander \at
              Dept.~of Physics \& Astronomy, University of Leicester, University Road, LE1 7RH, UK\\
              \email{richard.alexander@leicester.ac.uk}
}

\date{Received: date / Accepted: date}
% The correct dates will be entered by the editor

\maketitle

\begin{abstract}
We review the general theoretical concepts and observational constraints on the distribution and evolution of water vapor and ice in protoplanetary disks, with a focus on the Solar System.  Water is expected to freeze out at distances greater than 1-3 AU from solar-type central stars; more precise estimates are difficult to obtain due to uncertainties in the complex processes involved in disk evolution, including dust growth, settling, and radial drift, and the level of turbulence and viscous dissipation within disks.  Interferometric observations are now providing constraints on the positions of CO snow lines, but extrapolation to the unresolved regions where water ice sublimates will require much better theoretical understanding of mass and angular momentum transport in disks as well as more refined comparison of observations with sophisticated disk models.
% \PACS{PACS code1 \and PACS code2 \and more}
% \subclass{MSC code1 \and MSC code2 \and more}
\end{abstract}

\section{Introduction}

The first stages of planet formation begin within protoplanetary disks--the cloud of dust and
gas that circle young stars during their first few million years of evolution.  These disks form
at the same time as their central stars, as a molecular cloud core collapses under the force of gravity.  While much of the mass in the core eventually coalesces at the center to form a young star, the large amount of angular momentum in the core dictates that some of the mass settles into orbit around that star.  It is
within these disks that solid particles aggregate to form the seeds of planets and satellites, with the chemical compositions of these solids determined by the different environments that they are exposed to within the disk.

The physical and chemical conditions within protoplanetary disks evolve over millions of years as mass and angular momentum are transported during the final stages of their star's growth.  Their structures are 
determined by a complex interplay between dust evolution, which affects the disk temperature, 
density and ionization distributions through absorption of light from the central star
and any internal viscous heating, and transport processes which cause accretion, diffusion,
and possibly turbulence (each of which affect the dust distribution). In particular, the specific locations of water ice and vapor within the disk, and therefore the implications for the possible delivery of water to terrestrial planets, remain uncertain due to uncertainties in transport, viscous heating, and
dust growth and settling, as we discuss in this review.

\subsection{Protoplanetary Disk Basics}
The standard reference point in discussing the structure of protoplanetary disks is
the Minimum Mass Solar Nebula \citep{weid77,hayashi81}. The idea for this hypothetical 
disk came from trying to reconstruct the solar nebula
by taking the mass of planets in our Solar System and distributing it in a series of
"feeding zones" for each planet to recover a smooth distribution.  Hydrogen gas was
then added to recover a solar mix of elements (dust-to-gas ratio of $\sim$0.01 when water
ice was condensed and $\sim$0.005 when water remained as a vapor).  The total surface 
density (mass per unit area in a given annulus) of such a disk is given
by:
\begin{equation}
\Sigma \left( r \right) = 1700 \left( \frac{r}{\mathrm{1 ~AU}} \right)^{-\frac{3}{2}} \mathrm{~ g \, cm^{-2}}\,.
\label{eq:sigmammsn}
\end{equation}
This gives a total disk mass out to 100 AU of $\sim$0.01$M_{\odot}$.
The temperature structure was found by determining the directly-irradiated blackbody temperature as a function 
of distance from the Sun \citep{hayashi81}:
\begin{equation}
T \left( r \right) = 280 \left( \frac{r}{\mathrm{1 ~AU}} \right)^{-\frac{1}{2}} \mathrm{~ K}\,.
\label{eq:tmmsn}
\end{equation}
It is not clear how accurately the above structure
reflects the properties of the solar nebula. For example, there are significant uncertainties in
the temperature structures of protoplanetary disks
because they are optically thick, at least to the star's radiation and in parts to their own internal radiation field, and viscous dissipation may serve as an additional source of
heating, so that the application of equation (\ref{eq:tmmsn}) is questionable (see discussion in \S \ref{sec:t}).
As we discuss further below, disks are also evolving due to a variety of processes, transporting mass inward to be accreted by the central
star or losing mass due to photoevaporative radiation; thus the surface density and temperature are unlikely to be constant in time.
Moreover, the mass of the MMSN is a {\em minimum} by definition, and other models suggest higher surface densities \citep{desch07}. 

Despite these issues, the MMSN serves as a reference model to illustrate
key properties and structures by which we can understand real
protoplanetary disks.  Proceeding further, one must consider the vertical structure of the disk,
assumed to be set by hydrostatic equilibrium as follows: 
%In addition to the radial variations in disk properties given by
%Equations \ref{eq:sigmammsn} and \ref{eq:tmmsn},
%the vertical structure of the disk is also important.
%This structure is set by hydrostatic equilibrium, allowing
%the gas density to follow:
%
\begin{equation}
\rho_{g} \left( z \right) = \frac{ \Sigma}{\left( 2 \pi \right)^{1/2} h_{g}} \, \exp \left ( -\frac{z^{2}}{2 h_{g}^{2}} \right)\,,
\end{equation}
where $\rho_{g}$ is the $z$ is the height above/below the disk midplane and $h_{g}$ is the disk scale-height, 
with $h_{g}=c_{s}$/$\Omega$, where $c_{s}$ is the local speed of sound and $\Omega$ the local 
Keplerian frequency. For simplicity, the disk is assumed to be vertically isothermal, allowing $c_{s}$ and $h_{g}$ to be constant with height.  Figure 1 shows the plot of the gas density structure for the MMSN 
around the terrestrial planet region of the Solar System.  In the absence of any dust
dynamics (see below), small dust grains will be suspended within the disk, at least
initially, determined largely by whether water is condensed or not as described above.

\begin{figure}
    \centering
    \includegraphics[angle=90,width=0.49\textwidth]{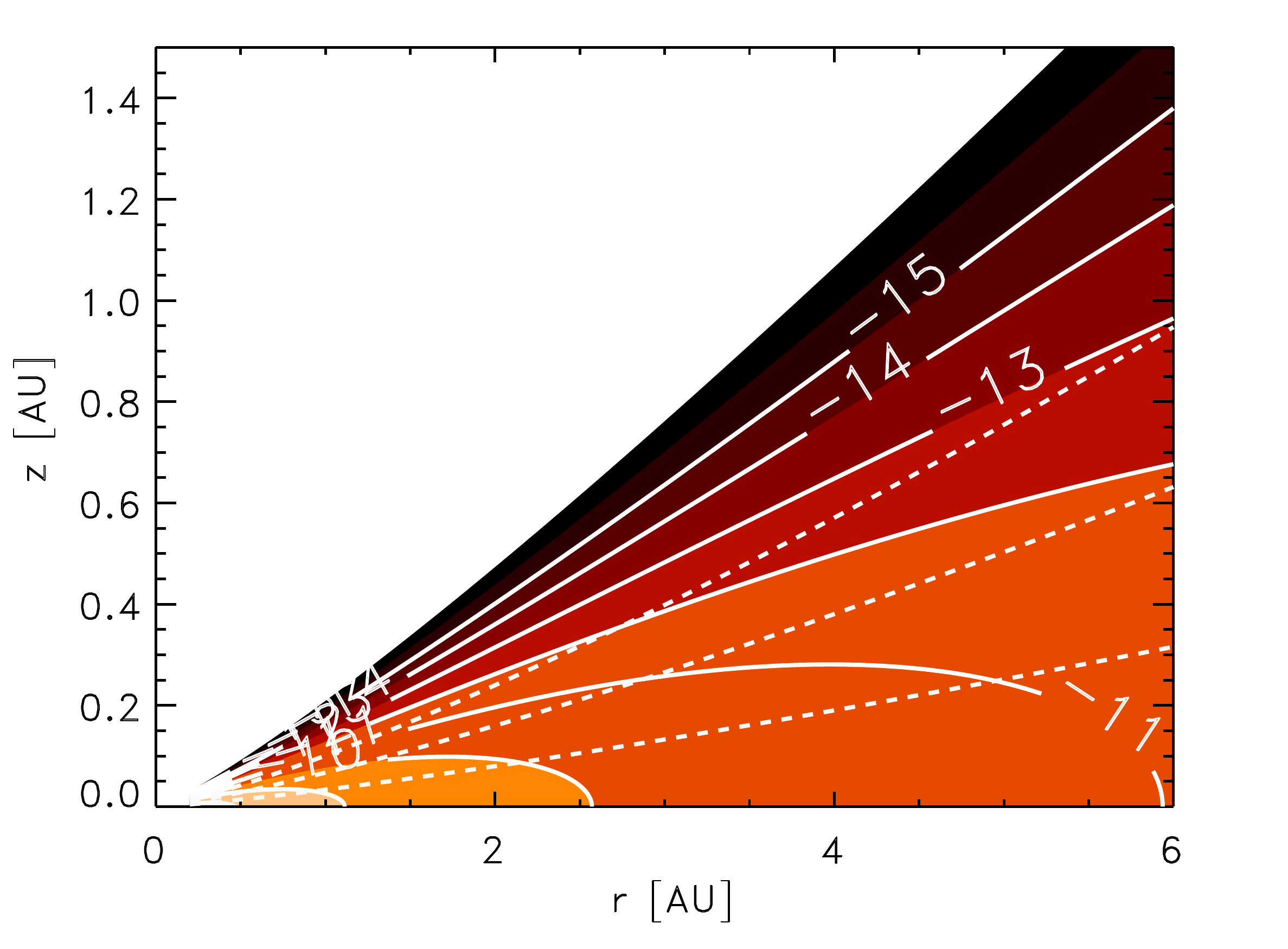}
    \includegraphics[angle=90,width=0.49\textwidth]{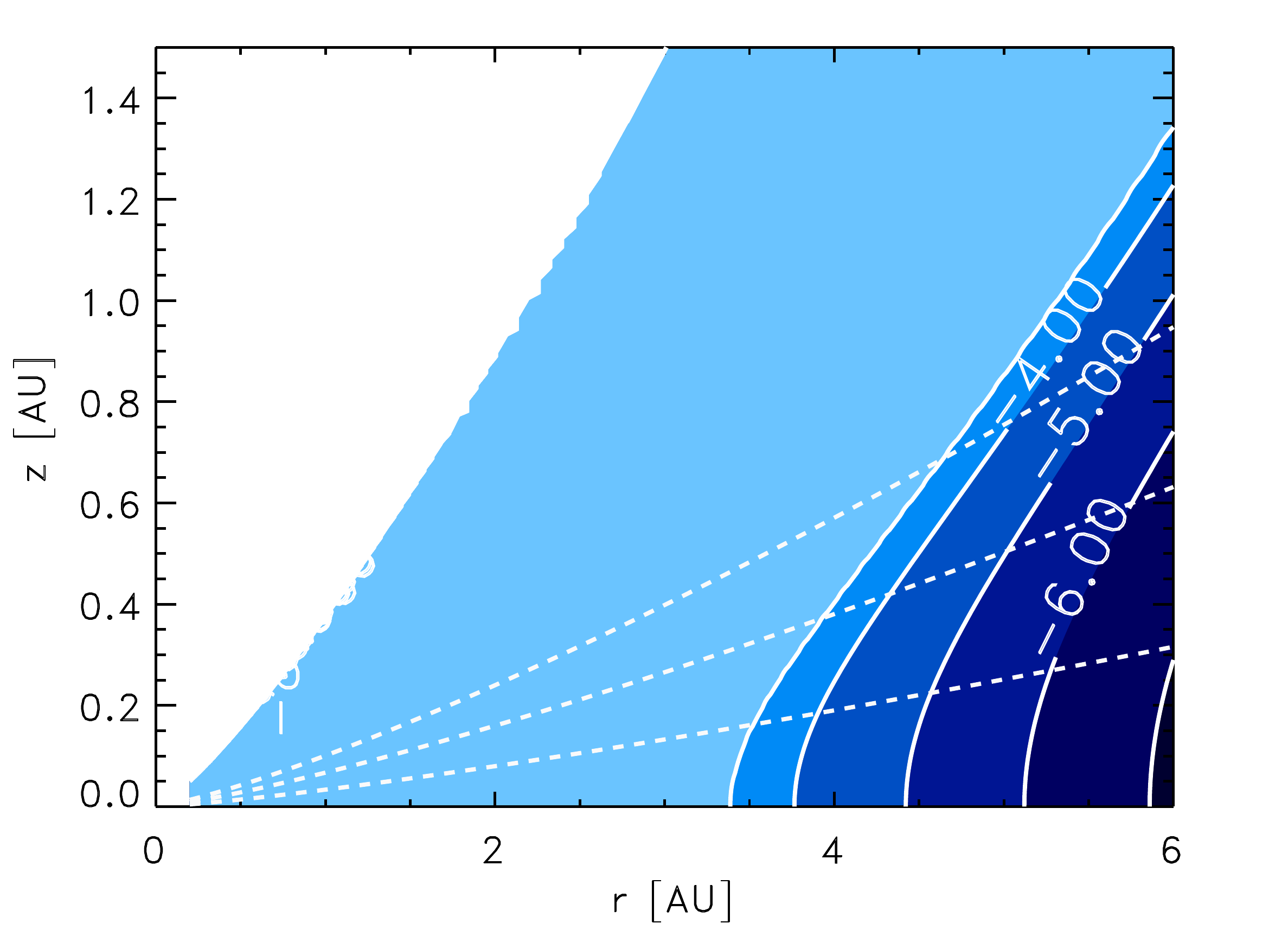}
    \caption{Left: Two-dimensional density structure of the inner regions of 
    the Minimum Mass Solar Nebula as described in the text.  Contours represent
    log$_{10}$ of the total gas mass density.  The dashed lines represent the
    heights of 1, 2, and 3 scale heights within the disk.  Right: Two-dimensional
    distribution of water vapor in the disk.  Canonical H$_{2}$O/H$_{2}$ number
    ratios are 5$\times$10$^{-4}$, which are found inwards of the midplane snow
    line (just inward of 3.5 AU here).  Contours are log$_{10}$ of the H$_{2}$O/H$_{2}$ ratio.}
\end{figure}

\subsection{Snow Line}

\label{sec:sl}
A key element in the discussion of water is the location of the snow line (SL) in the disk;
that is the distance from the star where water begins to freeze-out and become a solid.  Any water vapor that moves outward from this location condenses, while any water ice that
migrates inward evaporates, and thus this location is also
referred to as the water ice condensation/evaporation front.
This location can be determined from both equilibrium and
kinetic arguments.  In the equilibrium approach, the snow line is set based on thermodynamic data; that is, it is the location in the disk where, if all water was present as a
vapor in the disk, its partial pressure would exceed the
equilibrium vapor pressure at the local temperature.

An alternative approach defines the location of the snow line based on chemical kinetics, which is useful as thermodynamic equilibrium is not always achieved in a protoplanetary disk.  In this manner, the snow
line can be determined by finding where the freeze-out (adsorption)
flux of water molecules onto grains equals the desorption flux of 
molecules off of grain surfaces.  Following \citet{bergin13}, 
the two fluxes can be expressed as:
\begin{equation}
F_{\rm ads} = 0.25\, n_{H_{2}O}\, v_{th}\,,
\end{equation}
\begin{equation}
F_{\rm des} = \nu_{0}\, exp(-E_{b}/kT) N_{s} \,,
\end{equation}
where $n_{H_{2}O}$ is the local number density of gaseous water molecules, $v_{th}$ is the 
thermal velocity of those water molecules, $E_{b}$ is the experimentally determined 
binding energy of water onto a grain surface, $N_{s}$ is the density of adsorption sites 
on the surface of a grain ($\sim$10$^{15}$ cm$^{-2}$), and $\nu_{0}$ is a vibrational 
frequency of order 10$^{12}$ s$^{-1}$.  In instances where $F_{\rm des} > F_{\rm ads}$, water 
remains completely in the vapor phase.  When $F_{\rm ads} > F_{\rm des}$, water molecules begin
to freeze-out onto dust grains until $n_{H_{2}O}$ drops to a value such that the two
fluxes are equal.
Taking the 
 water (vapor+ice) to hydrogen
ratio throughout the disk as a constant $n_{H_{2}O}^{Tot}$/$n_{H_{2}} \sim 5\times 10^{-4}$ \citep{lodders03,cleeves14} and typical protoplanetary disk midplane ($z$=0) conditions, the snow line will generally occur
at temperatures of $T \sim$120-170 K.  These temperatures are in general agreement with those predicted from equilibrium studies \citep[e.g.][]{lodders03}.

The snow line is often pictured as being a singular radial location in the disk, $R_{\rm SL}$, 
the point at the disk midplane where $F_{\rm ads} = F_{\rm des}$ can only be maintained if some 
of the water in the vapor phase is removed and put onto solids. 
However, due to the 
decrease in gas density with height above the disk midplane, the snow line actually has 
a vertical structure.  If the disk is vertically isothermal, 
$F_{\rm ads}$ will decrease exponentially with height (following the background gas density as assumed in Eq. 3), 
while the desorption flux will remain constant.  Thus, the snow line will begin at the 
disk midplane at $R_{\rm SL}$, with only a small fraction of water being incorporated into 
solids and the rest remaining as vapor with height. As seen in Fig 1. for $r>R_{\rm SL}$, the amount of water in the disk present as ice increases; this is due both to the increasing height of the snow line and the drop of temperatures leading to more water being removed from the gas phase. 

\subsection{Dust and Gas Dynamics in Protoplanetary Disks}
\label{sec:dust_dynamics}
Upon formation of a protoplanetary disk, 
the dust grains and gas are generally thought to be uniformly mixed, with $\sim$99\% of the mass 
found in gas and $\sim$1\% as condensed solid grains ranging from submicron to microns 
in size \citep{lodders03}.  Over time, the solids and gas will be subjected to a series of dynamic processes 
which will lead to them evolving differently from one another and aid in the formation of 
planetary building blocks.

While the vertical structure of the gas in a protoplanetary disk is determined by the hydrostatic balance 
between the component of stellar gravity that pulls material towards the disk midplane 
with the pressure of the gas, solids feel less pressure support and will 
settle through the gas toward the midplane.  Larger solids settle more rapidly 
than fine dust.  Turbulence, if present, will work to offset the growing concentration 
of solids at the disk midplane, lofting materials to higher atltitudes again.  The extent 
to which solids of a given radius, $a$, settle in a disk is determined by the dimensionless 
stopping time of the particle:
\begin{equation}
    \tau_{s} = \frac{ \rho a}{\rho_{g} c_{s}} \Omega
\end{equation}
where $\rho$ is the material density of the solid. \citep[This quantity is also referred to as the Stokes number in the literature.  e.g.][]{cuzzi93}.
The amount of turbulence in a disk is uncertain, as described below. Models typically quantify it by defining a dimensionless parameter, $\alpha$ which produces an effective viscosity, $\nu$=$\alpha c_{s} h_{g}$.  This turbulence would also produce a diffusivity within the
gas which is often taken as $D$ as being of order $\nu$, though differences in magnitude are possible.  If we assume $D$=$\nu$, for the
particles we are interested in, then those with 
$\tau_{s}$/$\alpha <<$ 1 will remain suspended throughout the height of the disk 
(giving them a dust layer thickness of $h_{d} \sim h_{g}$), while larger particles will 
settle into a dust layer with thickness \citep{youdinlithwick07}:
\begin{equation}
h_{d} = \left[ 1 + \frac{\tau_{s}}{\alpha} \left( \frac{1+2 \tau_{s}}{1+ \tau_{s}} \right)\right ] ^{-\frac{1}{2}} h_{g} \,.
\end{equation}
It is worth noting that larger particles, as they decouple from the
gas, will have effective diffusivities less than that of the gas.
Turbulence may also redistribute materials radially in the disk, but this is only 
important for gases and small solids (those small enough to remain coupled to the gas 
with $\tau_{s} < \alpha$).  

Larger particles will also begin to decouple from the gas through their radial motions 
in the disk.  While the gas in the disk orbits its central star, its velocity is usually
slightly less than Keplerian due to the generally outward pressure gradient (hot, dense 
gas near the star; cool, sparse gas further out). Thus, while the star exerts a 
gravitational force on the gas, the net central force is somewhat less due to the pressure 
support it experiences.  Solid particles get negligible support from this pressure 
gradient, and thus would otherwise orbit the star at Keplerian rates.  As a result, the 
solids experience a headwind in their orbits around the star.  Solids with $\tau_{s}\!\ll$1 continue to remain 
coupled to the gas and are minimally affected by aerodynamic drag, while those with 
$\tau_{s} \gg$1 have so much inertia that the headwind can largely be ignored.  However, those 
particles with $\tau_{s} \sim$0.01-10 (typically mm to meter sizes) drift inwards as 
a result of the loss of energy and angular momentum arising due to the headwind.

The collective effect of these dynamical mechanisms results
in a complex evolution of the dust-to-gas ratio as a function of time, with important but uncertain consequences for the
building of large bodies and the global water distribution, as discussed next.

\subsection{Viscous heating and radiative transfer effects}
\label{sec:t}

The simple optically-thin equation (\ref{eq:tmmsn}) for the disk temperature is unrealistic because protoplanetary
disks are optically thick to radiation from the central
star.  Beyond a few AU or so, temperatures are
dominated by irradiation from the central star.
The precise heating rate is dominated by the ``flaring''
of the disk (i.e., the layers which absorb the starlight
lie at heights $z/R \propto R^{\eta}$, $\eta > 0$;
\cite{kenyon87}).  In the case of dust particles with properties
similar to that of the interstellar medium (ISM) that are
well-mixed with the gas,
the simple model of \citet{chiang97} for optically-thick disk absorption of starlight yields an
effective (disk surface) temperature
distribution $T_{e,r} \propto R^{-3/7}$, while more detailed models
yield a slightly steeper dependence on radius,
\begin{equation}
T_{e,r} \approx 220 \, L_1^{1/4} R_{AU}^{-1/2}\,{\rm K}\,
\end{equation}
\citep{dalessio98,dalessio01},
where the stellar luminosity is in units of
solar.  However, the {\em midplane} temperature for
such optically-thick disks is considerably lower,
often by a factor of two at
a few AU. It appears that
irradiation heating by itself is not able
to move the water snow line significantly
out beyond 1 AU, which poses problems for
understanding how chondritic meteorite parent bodies, thought to form outside of $\sim$1 AU late in solar nebula evolution, are water-poor \citep[e.g.][]{abe00}.

The apparent limitations of stellar irradiation
for explaining the variation of water abundances in the solar system has led to 
proposals that accretion heating is the
dominant factor setting the disk central temperature \citep[e.g.,][]{garaud07,kennedy08,bitsch15}.
To fix ideas, consider a toy model of
a optically-thick, steady, viscously accreting disk, with
effective temperature of
\begin{equation}
T_{e,a} = \left( {3 G M \mdot \over 8 \pi \sigma R^3} \right)^{\frac{1}{4}}
\sim 85 \, M_1\, \mdot_{-8} R_{AU}^{-3/4}\,{\rm K}\,,
\end{equation}
where we have assumed fiducial values of a central
star of one solar mass and an accretion rate of
$10^{-8} \msunyr$ as typically observed.
If the disk is very optically thick and the viscous dissipation occurs near the midplane, radiative
trapping can increase the central temperature
$T_{c,a}$ over $T_{e,a}$ by a factor of $\sim 
(3 \tau /8)^{1/4}$ \citep{hubeny90}, where 
$\tau$ is the vertical Rosseland mean
optical depth.  The optical depth is
$\tau = k_R \Sigma/2$, where
$k_R$ is the Rosseland mean opacity
and $\Sigma$ is the surface density of the disk, which
is related to the mass accretion rate $\mdot$ in the standard (vertically-isothermal) viscous disk as
\begin{equation}
\mdot = 3 \pi \alpha c_s^2 \Sigma \Omega^{-1}\,,
\end{equation}
bf where $\alpha$ is the usual parameterization of 
the viscosity (see \S \ref{sec:angmom}) and $c_s$ is the sound speed.  The result is
\begin{equation}
T_{c,a} = \left [ {3 \over 64 \pi^2}{G M \mdot^2 \over \sigma R^3} {\mu \Omega k_R \over \alpha R_g} \right ]^{1/5} 
= 330  \, M_1^{3/10}\, \mdot_{-8}^{2/5} k_R^{1/5}
\alpha_{-3}^{-1/5} R_{AU}^{-9/10}\, {\rm K}\,,
\label{eq:tvisc}
\end{equation}
where $\alpha_{-3}$ is in units of $10^{-3}$ and $R_{g}$ is the gas constant, and for simplicity we have ignored any
temperature or density dependence of $k_R$.

\begin{figure}
\centering
\includegraphics[width=0.8\textwidth]{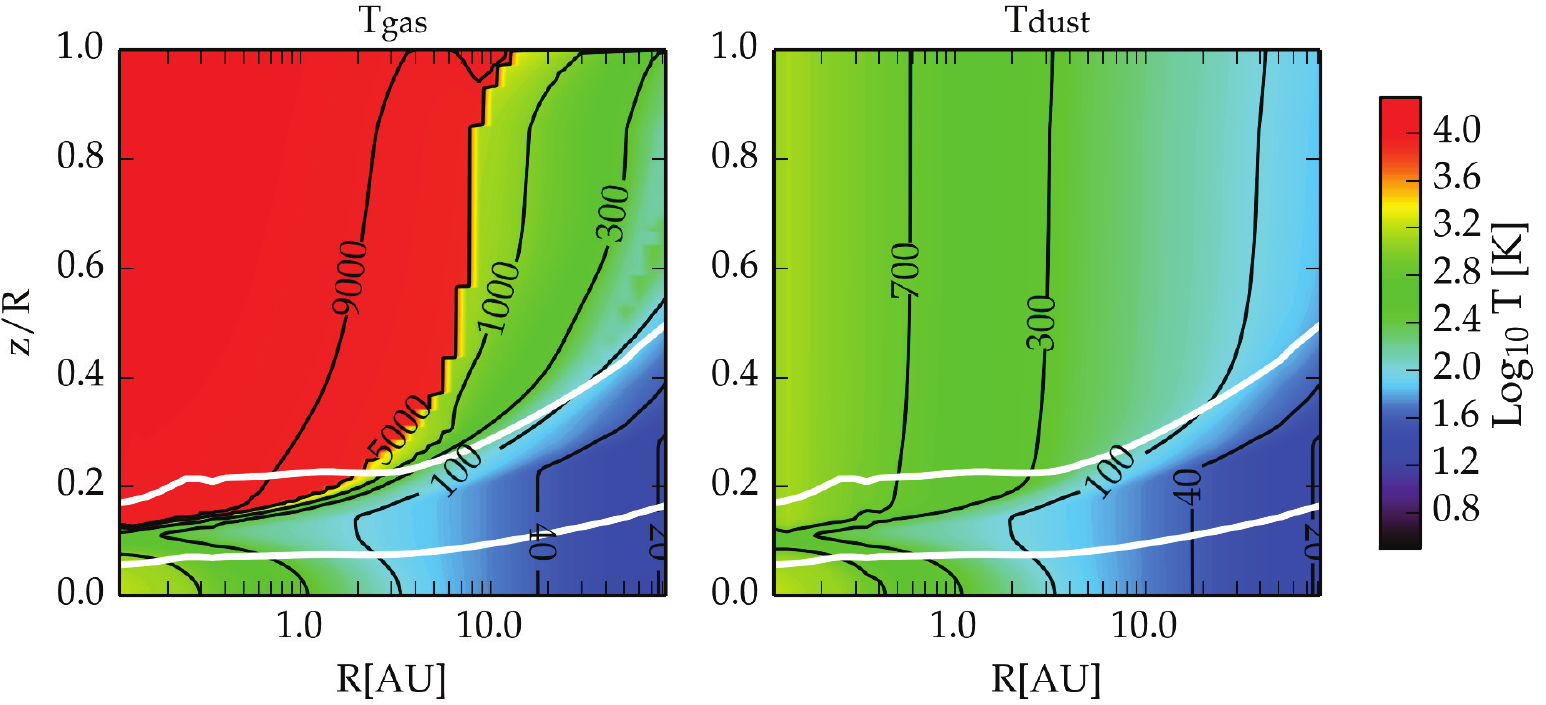}
\includegraphics[width=0.8\textwidth]{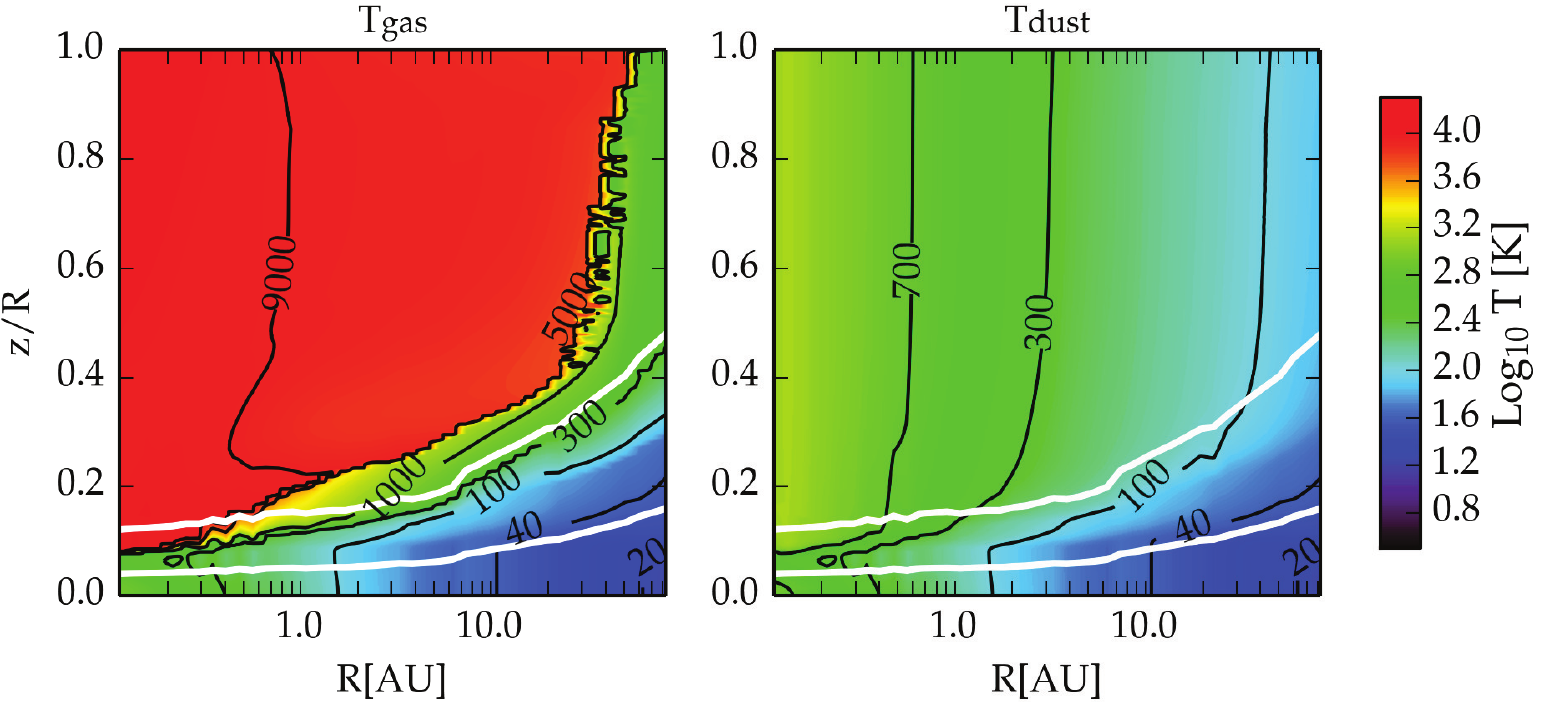}
\caption{Upper panels: gas (left) and dust (right) temperatures
for a detailed radiative transfer model with viscous
heating with $\mdot = 10^{-8} \msunyr$, $\alpha = 10^{-3}$.
The SL at the midplane lies at $R \sim 2-3$ AU,
in agreement with equation \ref{eq:tvisc} for the toy
model.
Lower panels: the same global model properties, but now
assuming that small dust is depleted by a factor of 100
in the upper layers.  Now the SL moves inward
to $\sim$ 1 AU, due to the decreased trapping of the
viscously-generated radiation.  Courtesy Nuria Calvet
and Ramiro Franco Hernandez.}
\label{fig:tgd}
\end{figure}

Assuming ice condenses at temperatures
170-120 K, and setting $k_R \sim 1 \,{\rm cm^2 ~g^{-1}}$, the crude estimate of equation \ref{eq:tvisc} indicates a snow line at $\sim 2.1-3.1$~AU for the fiducial parameters, is reasonably consistent with more
detailed calculations \citep[e.g.,][see Figure \ref{fig:tgd}]{bitsch15}.  However, this estimate may well be an upper limit to the increase in temperature due to viscous heating.
As discussed in
\S \ref{sec:dust_dynamics}, dust evolution is likely
to reduce the optical depth, as growth beyond the ISM grain size
distribution will have lower
opacities. Emprical estimates suggest depletion of small dust in the upper disk layers by 
factors of $10^{-1}$ to $10^{-3}$ \citep[\S 2.1,][]{furlan06}; presumably this is the result of growth as the large particles settle to the midplane \citep[as needed to explain mm emission;][]{ dalessio01}). 
The lower panels of Figure \ref{fig:tgd} show that
the snow line moves in to $R \sim 1$~AU for a depletion
of $10^{-2}$ of small dust in the upper layers.  Theory
suggests that such depletion should occur rapidly
(depending upon the level of turbulence present),
so the assumption of no depletion seems problematic.

Higher accretion rates and fragmentation to replenish small dust grains can help compensate for reduced
trapping of heat by dust depletion \citep[e.g.,][]{bitsch15}.   However, mass accretion rates drop significantly
with time, on timescales of 3 Myr or less
(\S \ref{sec:gas}; \cite{hartmann16}) and dust growth even with fragmentation is rapid
(\S \ref{sec:dust}). 
%Thus, relying
%on higher $\mdot$ to move the snow line out further
%implies rapid formation of solid bodies that later do
%not accrete much mass \citep{morbidelli15}.
Morever, it is not clear that viscous dissipation occurs mostly
near the midplane, or even that viscosity is important in transporting angular momentum in large regions of the disk, making its importance as
a heat source questionable (see \S \ref{sec:angmom}).
% - 
%could increase the surface density and radiative trapping,
%moving  the snow line out by a factor of $\sim 1.7$ relative
%to the fiducial values.  (It also implies a disk surface
%density much larger than that of the MMSN.)
%However, this assumes the angular momentum 
%transport is dominated by processes that heat the
%disk (e.g., via viscous dissipation).
%For example, disk wind models 
%drain most of the energy released by
%accretion into the wind, with essentially negligible
%dissipation in the disk interior (\S \ref{sec:non-turb}).

Major outbursts of rapid accretion - up to 
$10^{-4} \msunyr$ - are known to occur in some young objects (the FU Ori variables; \citealp{hartmann96}), whose disks
are demonstrably internally-heated and thus
viscous-dissipation dominated.  Such high accretion
rates can move snow lines farther out \citep{cieza16}, but
these large outbursts are confined to early evolutionary phases
(ages $< 1$~Myr); any remnants of these episodes may well be
accreted into the star by the time of planetesimal formation.

\section{Observational constraints on disk structure and evolution }
\label{sec:obsdiskev}

The density and internal temperature
structure of protoplanetary disks are not well
constrained.  Most of our direct information
is limited to
scales of several to to hundreds of AU.  The
inner disk structure near the expected snow line at a few
AU (for a solar-luminosity star) is not easily resolvable
with current techniques, and there are indications that inner
disks are optically thick (due to dust) at mm wavelengths. One therefore
must work to connect models and observations
on larger scales to attempt any extrapolation to
the water SL.

\subsection{Dust}
\label{sec:dust}

\begin{figure}
    \centering
    \includegraphics[width=0.45\textwidth]{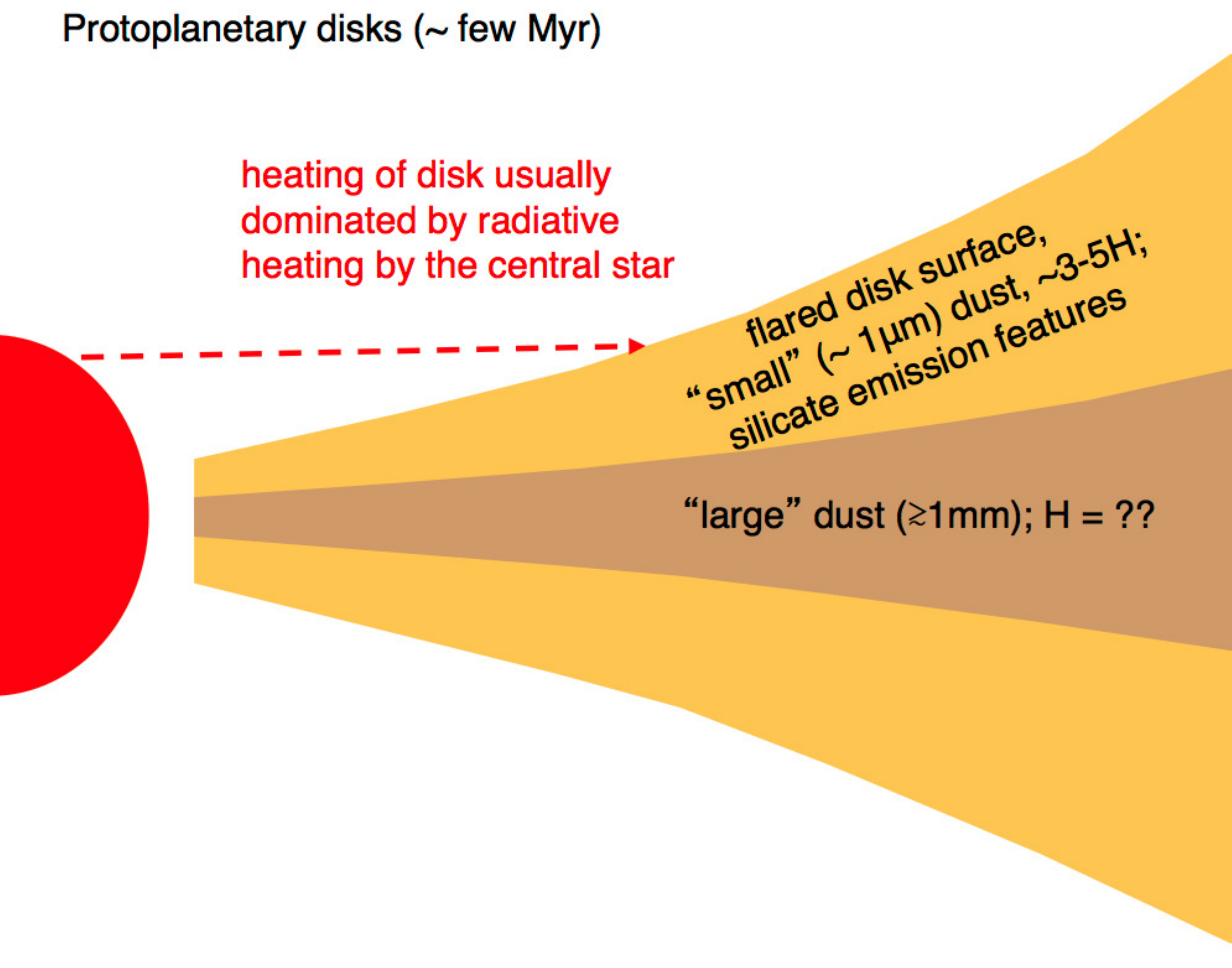}
    \includegraphics[width=0.45\textwidth]{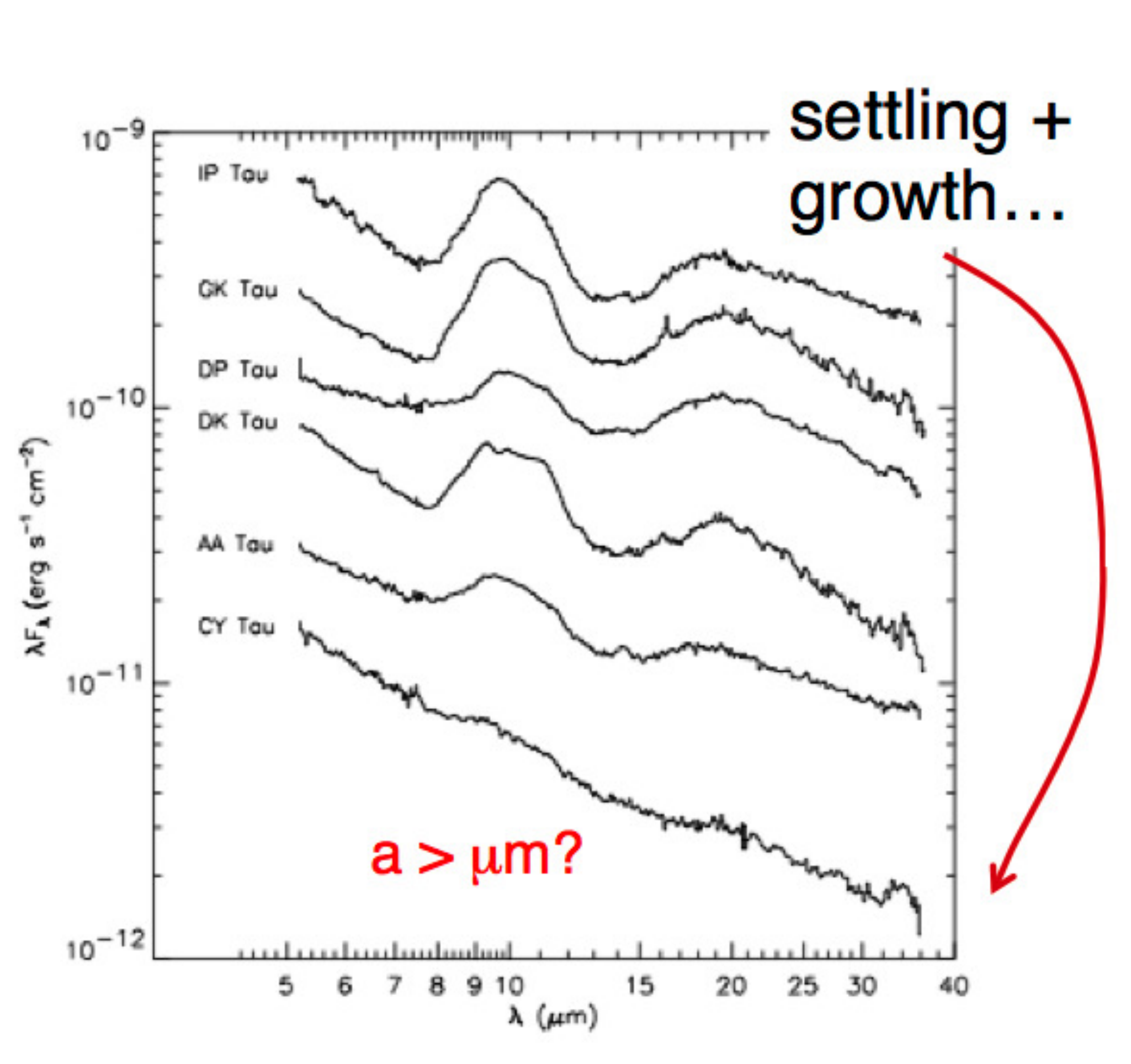}
    \caption{The left panel schematically indicates structure of solar-mass pre-main sequence (T Tauri) disks.  Small dust must be mixed with the gas to at least a few scale heights to explain the observed infrared dust emission (right panel), both in terms of a continuum and also to explain the silicate emission feature at $\lambda \sim 10 \mu$m. Large (mm) dust must be present to explain the long-wavelength emission; presumably this dust resides in a relatively thin layer at the disk midplane.  The set of mid-infrared spectral
    energy distributions in the right panel, taken from \citet{furlan06}, is suggestive of a sequence of increased settling of dust, making the infrared excess fall off faster to longer wavelengths, along with a decrease in the strength of the silicate emission feature which may indicate grain growth. Note that while some stars exhibit emission consistent with amorphous silicates (e.g., IP Tau), others clearly show the presence of crystalline silicate particles (e.g., DK Tau). }
    \label{fig:sedscheme}
\end{figure}

Most of what we infer about the radial and vertical temperature
and density distributions of disks
comes from observations of dust emission,
as cold molecular hydrogen is not detectable and most other
molecular tracers are either optically-thick, are activated only in
disk upper layers, or are
subject to uncertainties in chemical reaction rates. 
As shown in Figure \ref{fig:sedscheme}, we infer that the dust in disks around
solar-mass pre-main sequence (T Tauri) stars exhibit a sort of ``two layer'' structure, with large dust concentrated near the
disk midplane and small dust suspended in upper layers
to a few scale heights.  The $10-20 \mu$m silicate emission features 
seen in most T Tauri disks can be explained by dust with sizes smaller 
than a few $\mu$m, while the amount of mm-wave emission is only explainable with reasonable dust masses if the grains have sizes of order 1 mm or larger \citep[e.g.,][]{dalessio01}.  Placing the small dust in upper layers where
the vertical temperature inversion (caused by irradiation due to the
central star) results in emission features, while distributing the
large dust closer to or in the midplane avoids the elimination of
silicate features.

As dust grows and settles to the midplane
(see \S\ref{sec:dust_dynamics}), small dust becomes depleted in the upper layers.  Indeed, theoretical
models of this process in the absence of significant
turbulence suggest this should happen
quickly, posing challenges to explain the observed
existence of suspended small dust
in disks of ages as small as 
$\sim 1$~Myr \citep{dullemond05, birnstiel11}.
Empirical estimates of small dust depletion in upper
disk layers are difficult to make with any accuracy,
but current models to explain observations suggest
depletions by factors of 10 to 1000 \citep{dalessio01,furlan06}, with
significant implications for radiative trapping of viscous heat
(\S \ref{sec:t}).

Efforts to demonstrate grain growth generally are based on observations of spectral indices in the mm-wave region.  
Small ISM dust
has a steep dependence of opacity on frequency
$k_{\nu} \propto \nu^{-\beta}$, with $\beta \sim 1.7$, as a
consequence of the limited sizes of the largest grains ($\sim 0.3\, \mu$). Observations of mm-wave and cm-wave
emission from disks at distances of $\sim 20-100$~AU indicate
much smaller values of $\beta \sim 1$ or even 0, indicating growth to at least mm or even cm sizes
\citep{perez12,perez15,tazzari15}. 
Unfortunately, maximum particle sizes cannot be determined once dust grows in
size beyond the wavelengths of observation; in this case $\beta$ is 
determined by the shape of the particle size distribution.
As emission at cm wavelengths
is generally quite weak in outer
disks (50-100 AU), currently it is not possible say anything about
$\simgreat$ cm-sized dust at large radii.

Given the uncertainties in dust (and solid) size
distributions, dust disk masses must be considered to
be lower limits; that is, what is being measured at mm
wavelengths is only the amount of mass in solids of sizes
$\sim 100 \mu$m to $\sim$~a few mm.  The survey of 
\citep{andrews05} yielded a median ``disk mass''
(100 x the solid mass) of $\sim 0.005 \msun$. Statistical
estimates from gas accretion yield somewhat larger values
(\S \ref{sec:gas}).

Another complicating effect in addition to vertical settling
is radial drift.  This effect has been seen in 
a number of objects, most clearly in the nearby T~Tauri
star TW Hya with mm-dust grains having a sharp edge at $\sim$60 AU, while CO
gas emission is seen out to $\sim$200 AU \citep{andrews12}. As
drift of mm--cm sized particles is quite rapid, this causes
some problems in understanding the existence of dust in
this size range out at $\simgreat 50$~AU, particularly in
this relatively old object (age $\sim 10$~Myr).

\subsection{Gas}

\label{sec:gas}
The only current constraint on the mass of molecular hydrogen
in disks is the estimate of $M_d \sim 0.06 \msun$ from
analysis of HD in TW Hya \citep{bergin13}.  Even this result
is model-dependent, as it relies both on the assumed
isotopic ratio and, more importantly, the disk temperature
structure, as HD does not probe the coldest regions where most of the mass resides.

\citet{williams14} attempted to use rotational emission lines of $^{12}$CO,
$^{13}$CO, and C$^{18}$O isotopologues to correct
for optical depth effects and variations in gas
temperatures as a function of radius and height.  Assuming a conversion of CO to H$_2$,
they derived gas masses of order
$10^{-3} \msun$,
an order of magnitude smaller than expected from dust masses
(times 100).  More importantly,
such gas masses are also an order of magnitude too small to sustain typical accretion rates of $10^{-8} \msunyr$ over timescales of 1-3 Myr (see below). One possible
resolution is that CO
freezes out onto grains, removing it from the gas phase and thus
rendering it undetectable in mm-wave rotational line emission.

\begin{figure}
    \centering
    \includegraphics[width=0.7\textwidth]{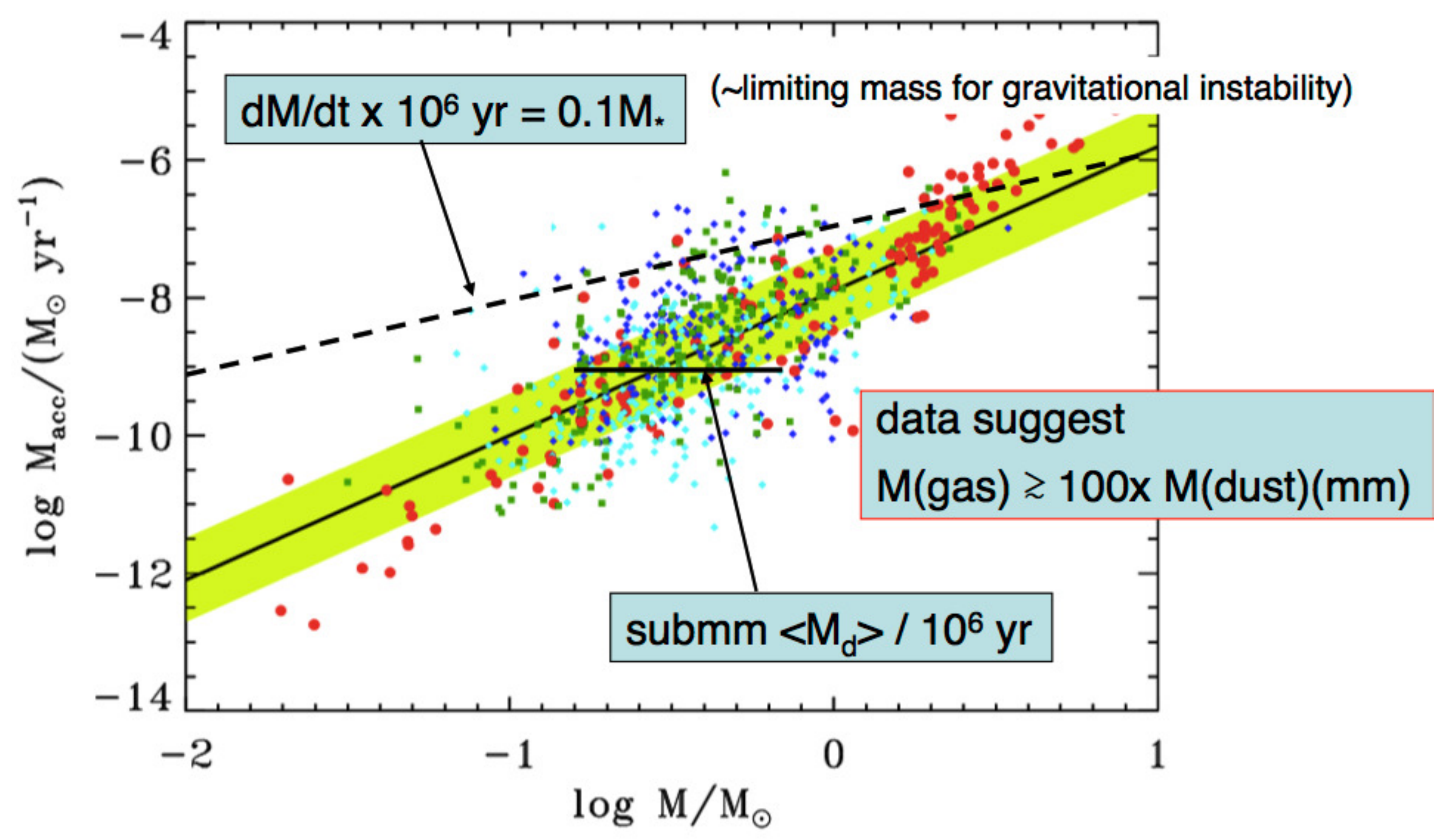}
    \caption{Observationally estimated accretion rates
    of T Tauri stars at typical ages $\sim 1-2$~Myr as a function of stellar mass.  There is a clear trend of increased mass accretion with increasing stellar mass, though there is substantial scatter.  The dashed line indicates the rate that would result in accreting 10\% of the stellar mass from the disk; it seems unlikely that disks can be considerably more massive than this, as they would be gravitationally
    unstable.  (Stars with masses $> 1 \msun$ appear to
    have shorter average disk lifetimes.)  The data suggest that T Tauri stars typically accrete a mass equal to the inferred dust masses, suggesting that the mm-wave emission analysis generally underestimates the gas mass. Modified from \citet{hartmann16}.}
    \label{fig:dmdt}
\end{figure}

\begin{figure}
    \centering
    \includegraphics[width=0.85\textwidth]{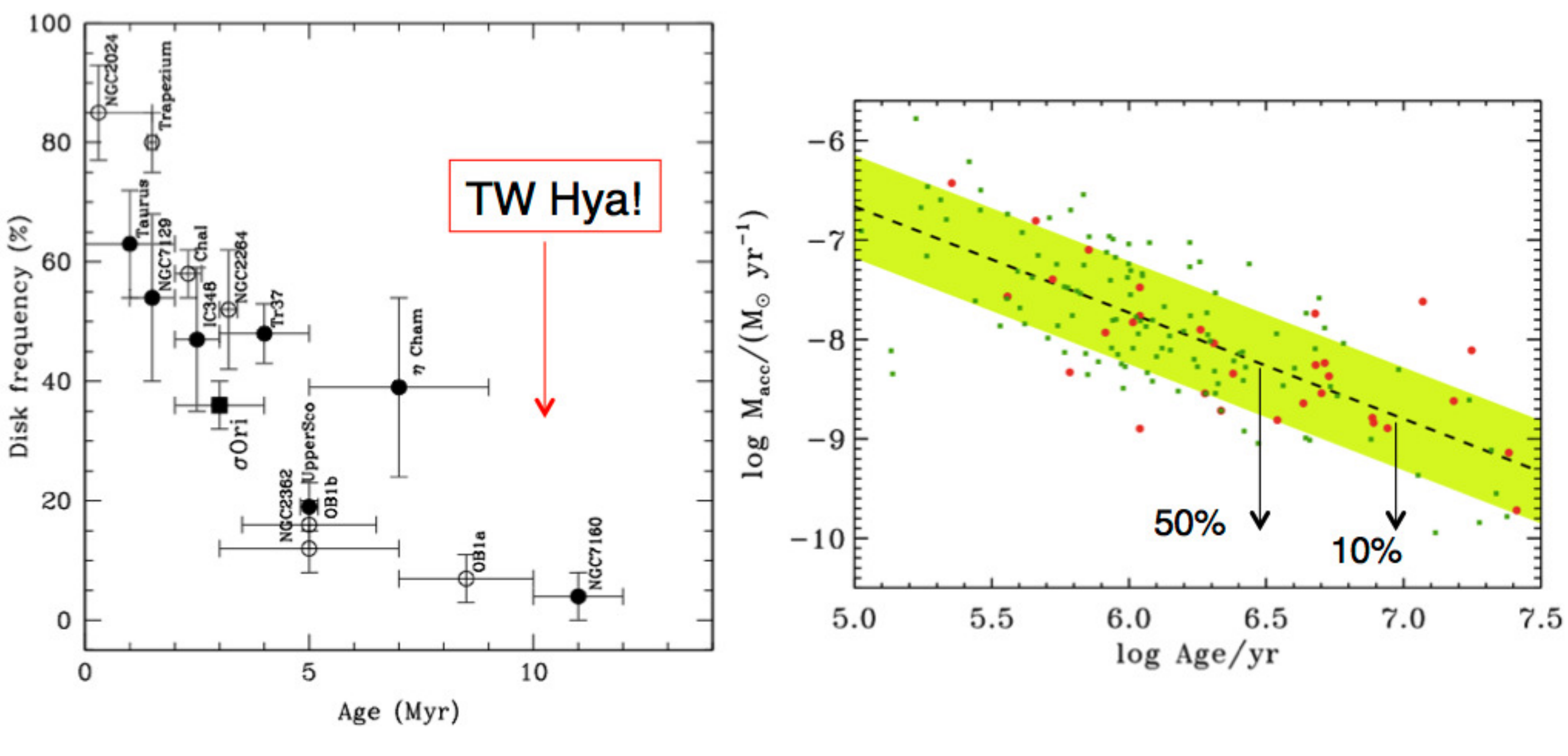}
    \caption{Left: frequency of detectable infrared dust disk emission in the IRAC bands 
    $3.5 - 8 \mu$m as a function of group or cluster age.
    From \citet{hernandez08}.  Right: dependence of mass accretion rates on age, scaled to account for the mass dependence of equation \ref{eq:mdotm} for stars with masses in the range 
    0.3-1.0 $\msun$ (see also Fig.~\ref{fig:dmdt}).  The mass-accretion rate relationship is calculated for objects with mass between $0.05-1$ M$_\odot$. The resulting fit is uncertain due to correlated errors (see text).  The arrows emphasize that significant numbers of accreting disks ``drop out'' with increasing age.  From \citet{hartmann16}.}
    \label{fig:lifetimes}
\end{figure}

Gas mass accretion rates ($\mdot$) are
important in providing statistical upper limits to the possible viscous heating in disks.  $\mdot$ increases with increased stellar mass and decay with age.  The
best fit for T Tauri stars with a median age of about
1-2 Myr derived by G. Herczeg is
\begin{equation}
\log (\mdot /\msunyr)  \approx -7.9 +2.1 \times \log (M_*/\msun)\,,
\label{eq:mdotm}
\end{equation}
\citep[see][]{hartmann16}.  As one can see from Figure \ref{fig:dmdt}, there is a large scatter at any
mass, more than the factor $\sim 0.5$~dex expected for
observational errors. 
Accretion rates also generally decrease with increasing age (Figure \ref{fig:lifetimes}), although the rates of decay are uncertain, due in part to difficulties in assigning 
ages to individual young stars \citep{soderblom14}.
This decrease in $\mdot$ with time, along with the likely
depletion of small dust with increasing age, indicates
that the snow line should move inward during the
main optically thick disk lifetime (\S \ref{sec:t}).

The ``lifetimes'' of disks as judged from either infrared dust emission (left panel of Figure \ref{fig:lifetimes})
or from the ending of detectable accretion onto the
central star \citep{fedele10}, exhibit a significant
spread.  The ``usual'' disk lifetime is often
given as 3 Myr, but a few (optically-thick) disks
survive to $\sim 10$~Myr, while others last for
$\simless 2$~Myr.
It appears that both dust emission and accretion disappear
fairly rapidly (see below), so there must be something like a two-phase behavior, with a moderate decay lasting for 1-10 Myr and a rapid decrease over $< 1$~Myr.  This rapid clearing
could be produced by photoevaporative mass loss from the
disk driven by X-ray or other high-energy radiation
\citep{2006MNRAS.369..216A,owen11}, or by giant planet
formation.

\subsection{Gaps and holes}

\begin{figure}
    \centering
    \includegraphics[width=0.9\textwidth]{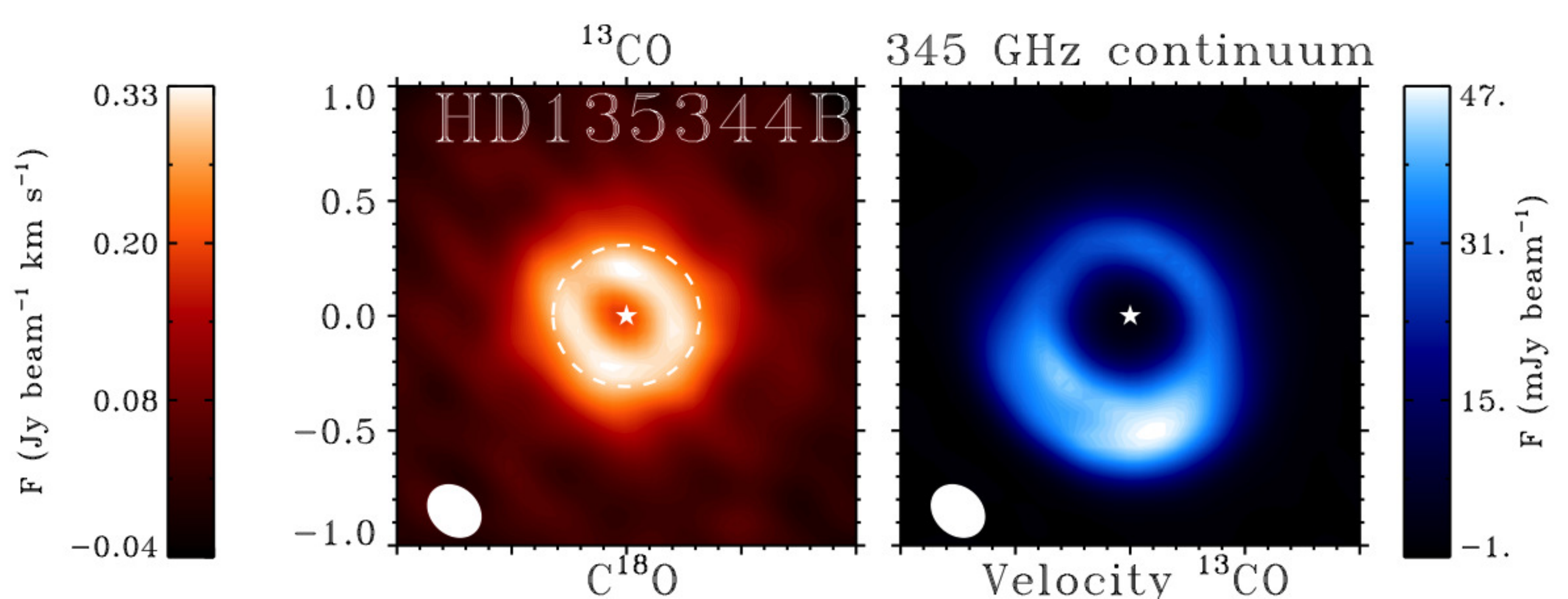}
    \caption{ALMA imaging of the $1.6 \msun$, $8 \lsun$
    young star HD 135344B, showing that gas emission (C$^{18}$O) is present inside the dust disk as seen in mm-wave emission. Similar structures are seen in several other systems, suggesting that planet formation has trapped large dust particles in the outer disk while allowing some gas to move inward, helping to explain the observed gas accretion onto central stars \citep[e.g.,][]{ingleby14}.  The non-axisymmetric
    structure in the dust may indicate a vortex, suggesting
    perturbations by planets.  From \citet{vandermarel16}.}
    \label{fig:vandermarel}
\end{figure}

The advent of sensitive, high-spatial resolution
mm-wave interferometry, especially using the
Atacama Large Millimeter and Submillimeter Array
(ALMA), has revealed that major radial
structuring of protoplanetary disks is far more common than
previously realized.  Disks of substantial masses but with
inner holes strongly depleted 
in small dust have been increasingly recognized (the ``transitional disks''; \citealp{calvet02,espaillat14};
Figure \ref{fig:vandermarel}).
In addition, there
are also disks with inner disk dust as well as
a large gap (the ``pre-transitional disks''; \citealp{espaillat07}).  Recent observations suggest
that the gas within these holes and gaps
\citep{bergin03,najita07,ingleby09} 
is also depleted, although by smaller factors than
of the dust \citep{vandermarel16}.  Because observations still generally
detect only the largest holes and gaps, the new
results suggest that there may not
be any truly ``primordial'' or ``protoplanetary'' disks once the infall of envelope material to the disk during the protostellar phase has ended.
Significant formation of planetesimals and embryos,
if not planetary cores, probably begins very early, $\simless$ 1 Myr,
especially if giant planets are responsible for
creating large disk holes and gaps.
This is consistent with evidence from the Solar System;
rapid formation of planetesimals is supported by radiometric dating
of iron meteorites, which suggest planetesimals formed in
our solar nebula concurrently or within ~10$^{5}$ years of
the formation of CAIs, the oldest objects in our 
Solar System, and which define $t$=0 in cosmochemical clocks
\citep[][]{kruijer14}.  Such bodies would predate the 
accretion of chondritic meteorite parent bodies by 2-3 Myr \citep{connelly12}.

An interesting sidelight to the presence of gaps and
holes strongly cleared of small dust in young disks
is the possibility that even rapid planetesimal and
embryo formation might occur in optically-thin regions
(see \S \ref{sec:final}).

\subsection{Observations of Water in Protoplanetary Disks}

Water vapor has been detected in several T Tauri
and Herbig Ae/Be stars \citep{carr11,brittain15,blevins15}.  H$_{2}$O
emission lines are seen superposed on the strong
dust continuum, showing that they arise in
upper layers of the disk, with typical gas temperatures
$\sim 600$~K \citep{carr11}.  Unfortunately, translating these observations of surface
water vapor, estimated to originate at
radii $\sim 0.5 - 5$~AU, into midplane snowline
positions is highly model-dependent
\citep{blevins15}, particularly
because the high temperatures of the upper disk
gas layers tend to reflect heating by stellar X-rays and ultraviolet radiation \citep{glassgold09}.

Detection of where water starts to freeze out onto grains has proven to be more challenging.
Even with the vast improvements made by the ALMA interferometry, spatial resolution is limited to a few AU in the nearest star-forming regions, and to the brightest objects.   As a check on our
general understanding of disk temperature structures,
 CO snow lines have been detected indirectly at scales of tens to nearly 100 AU \citep{qi13}, depending
upon the luminosity of the central star.  CO was not directly observed; instead, the inner edge of emission from N$_{2}$H$^{+}$ was taken to be the position of the CO snow line because this ion is only abundant in regions where CO is depleted from the gas.  Similar tracers do not exist for H$_{2}$O, making detection of the water snow line a challenge.  Further,
the likely positions of water snow lines in the nearest protoplanetary
disks are at the edge or below current imaging capabilities.
\citet{banzatti15} suggested that the water snow line might
be inferred from a change in the spectral index due to
the decrease in dust size and optical properties as water
ice is sublimated, but the potentially optically-thick
nature of the inner disks may make these observations difficult. 
Crystalline water ice has been detected in a
few disks from spatially-unresolved observations \citep{mcclure15}, but these observations undoubtedly
trace outer regions at tens of AU or more.

\begin{figure}
    \centering
    \includegraphics[width=0.8\textwidth]{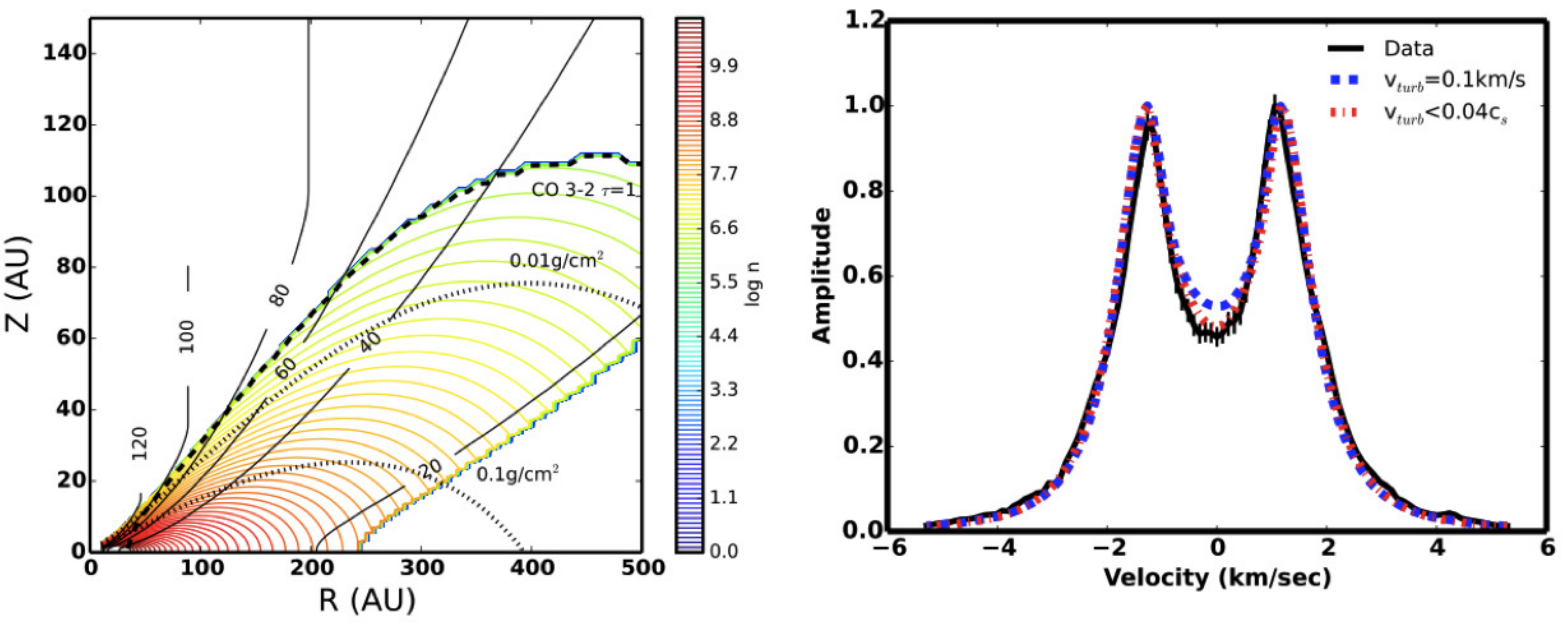}
    \caption{Left: disk model constructed to explain the
    CO 3-2 emission from the disk of HD 163296. The black solid curves denote isotherms in K, the dotted curves
    show contours of constant vertical column density, and
    the colored curves show isodensity contours for the
    CO lower level, with the upper heavy dashed curve showing
    the layer where the line becomes optically thick.  (The lower right region shows where CO freezes out.  This model produces the emission line profile shown in the right panel; the sharpness of the peaks and overall profile shape (modeled spatially) provides an indication that any turbulent broadening must be small.  From \citet{flaherty15}. }
    \label{fig:turb}
\end{figure}

\subsection{Observational limits on turbulence}
\label{sec:turbobs}

The magnitude of turbulence in planet-forming disks plays an important role in allowing (or preventing) coagulation/growth of solids, as
well as affecting both mass and angular momentum transport.
Unfortunately observational constraints on turbulence are
extremely limited.  Observations of the first-overtone
lines of CO have yielded estimates of transsonic turbulence in the upper layers of disks at radii $\simless 1$~AU in a few systems, but these objects are either relatively luminous \citep{najita96,najita09} or very rapidly accreting \citep{hartmann04}, such that temperatures are likely 
high enough to support the magnetorotational instability (MRI) via thermal ionization
(\S \ref{sec:angmom}). The most detailed limits so far
for colder disk regions
come from the recent ALMA study of the large disk surrounding the $L \sim 35 \lsun$ A-type star HD 163296 \citep{flaherty15}.
Using a sequence of CO isotopologues to deal with issues
of temperature structure, excitation, and optical depth,
Flaherty et al. conclude that non-thermal motions in the
upper disk layers are no more than about 3-6\% of the local
sound speed at $R \simgreat 30$~AU (see Figure \ref{fig:turb}), an order of magnitude less than expected
for full MRI turbulence, though still significant enough to play a role in mixing and transport of disk materials.

% ------------------------------------------------------------

%\section{The role of magnetic fields}
%\label{sec:magnetic}

\section{Turbulence, winds, and mass and angular momentum transport}
\label{sec:angmom}

Turbulence can play an important role in disk density and temperature structures via its effects on viscous dissipation and mass transport.  As discussed in 
\S \ref{sec:gas}, to maintain observed accretion rates over evolutionary timescales seems to imply transport of a significant fraction of initial disk masses into the central star.  
Turbulence can also
affect the amount of heating by light from the central star if it
stirrs up small dust grains \citep{dullemond05}, enhancing their vertical 
scale height relative to the gas \citep{2006A&A...452..751F}.

In the case of fully-ionized disks, the magneto-rotational instability
\citep[MRI:][]{balbus98} produces efficient turbulent angular momentum transport, with viscosities parameterized by the
parameter $\alpha \sim 10^{-1} - 10^{-2}$ (\S \ref{sec:t}). 
However, as they are comparatively dense and cold, protoplanetary disks (PPDs) are generally expected to have extremely low levels of ionisation deep in their interiors \citep[see][for a recent review]{2014prpl.conf..411T}.
Only inner ($\simless 0.3$~AU) regions of protoplanetary disks with higher temperatures and surface 
layers ionised by hard radiation (such as FUV photons, X-rays, cosmic rays) from the magnetically active central star \citep[see section 2.5 in][]{2014prpl.conf..411T}, or
potentially from an actively star-forming nearby cluster environment \citep{2015ApJ...815..112K},
can result in effective coupling
of the ions with the neutral gas.
Cosmic rays could be especially important in ionising gas, particularly in the outer disk where column densities are low, but deflection
 by the stellar wind or magnetic fields might prevent them from becoming important \citep{2013ApJ...772....5C}.

The ionising events from external sources are balanced by recombination with gas-phase ions and, in the presence of sub-micron-sized dust, by absorption on the grain surface.
Based on estimates of the abundance of free electrons and the collision regimes characteristic of various regions within the disc, it is now commonly accepted that all three non-ideal MHD effects, namely, Ohmic resistivity, ambipolar diffusion (AD), and the Hall term have to be considered in the generalised Ohm's law to arrive at sound conclusions for the overall dynamical state of PPDs \citep[e.g.][]{2003MNRAS.345..992S,2011ApJ...739...50B}.

%\subsection{Non-turbulent transport}

\subsection{Ambipolar diffusion affected laminar disks}
\label{sec:ad}

The inclusion of ambipolar diffusion has strongly modified the original layered model assuming
only Ohmic resistivity,
with an MRI-active upper disk layers providing the angular momentum transport and viscous heating \citep{1996ApJ...457..355G}.
assumed only Ohmic resistivity, with Studies by \citet{2013ApJ...769...76B}, \citet{2013ApJ...772...96B}, and \citet{2013ApJ...764...66S,2013ApJ...775...73S} show that
in the absence of net vertical flux, the MRI is typically suppressed in wide radial stretches of typical T~Tauri disks and standard solar-nebular models, and the resulting state is a laminar flow with negligible angular momentum transport.
Moreover, otherwise laminar regions in the inner disk (roughly speaking $R\simeq 1-20\,{\rm au}$) can launch a magnetocentrifugal wind within the FUV-ionised surface layers.  Such AD-assisted magnetic winds were actually explored in prescient work by \citet{konigl89} and \citet{wardle93}, who developed self-simular steady disk solutions matching to the wind solutions of \citet{bp82}, and illustrate many of the main features of subsequent numerical simulations \citet{2013ApJ...769...76B,2013ApJ...764...66S,2014A&A...566A..56L}.
However, due to technical limitations
(such as finite computational volumes), mass loading
and mass loss rates are uncertain.
Since the role of the far-UV ionised upper disk surface is crucial \citep{2011ApJ...735....8P}, the shielding of the required energetic photons needs to be assessed to corroborate the foundation of the mechanism.

A noteworthy feature of AD laminar 
accretion disks is the existence of thin current layers, which have both been found in the local models of \citet{2013ApJ...769...76B}, and the global ones of \citet{2015ApJ...801...84G}.
These current layers can produce strong radial streams of material (at $\sim30-40$\% of the sound speed), and can potentially provide non-negligible amounts of radial mass transport.
As has recently been argued by \citet{2016ApJ...818..152B}, laminar MHD winds can in principle team up with photoevaporative acceleration \citep[e.g.][]{2006MNRAS.369..216A} to remove mass and limit disk
lifetimes (see 
\S \ref{sec:photoevap}).
%Because winds launched away from the disk midplane preferentially remove the less dusty surface layers \citep{2016ApJ...821...80B}, they enrich the disk body in solids, in agreement with recent ALMA observations by \citet{{2016arXiv160405719A}}, who found dust enhancement in a range of disks in Lupus.

\subsection{Magnetic self-organisation and the Hall effect}
\label{sec:hall}

Although suggestions of the importance of the Hall effect in PPDs
have been made for some time \citep[e.g.,][]{wardle93}, it has only
recently been included in detailed numerical simulations.  Complicating
matters, the dynamical outcome of the Hall effect depends upon
the mutual sign of the rotation vector (represented by $\mathbf{\Omega}$) and large-scale magnetic flux (represented by $\mathbf{B}$).
The Hall drift 
has the effect to either boost (in the case of $\mathbf{\Omega}\cdot\mathbf{B}>0$) or impede (conversely for $\mathbf{\Omega}\cdot\mathbf{B}<0$) linear growth \citep{2014A&A...566A..56L,2015ApJ...798...84B} of the MRI.
Notably, with a somewhat modified ionisation model, \citet{2015MNRAS.454.1117S} found bursty behaviour in the latter case.

While the MRI relies on the combination of field induction and plasma inertia (described by perturbations $\delta\mathbf{v}$ and $\delta\mathbf{B}$), there is an even simpler instability that strictly only relies on the induction equation (and perturbations $\delta\mathbf{B}$) --
this has been termed the Hall-shear instability \citep[HSI,][]{2008MNRAS.385.1494K}.
As studied for the case of a simplified Hall coefficient that results in a fixed length scale, $l_{\rm H}\equiv \eta_{\rm H}/v_{\rm A}$, the HSI can lead to self-organisation behaviour (for $l_{\rm H}\simeq H$) and, depending on the strength of the effect, creates azimuthally confined zonal bands and/or vortices.

MRI in the context of Hall-MHD has so far been studied in unstratified \citep{2002ApJ...577..534S,2013MNRAS.434.2295K} and stratified \citep{2014A&A...566A..56L,2014ApJ...791..137B} shearing-box models, and in  unstratified cylindrical disk models \citep{2014MNRAS.441..571O,2016A&A...589A..87B}.
The microphysics of the Hall effect is complicated by many uncertainties,
such as the population of small grains and their charge.

The presence of strong self-organising azimuthal field belts naturally begs the question of buoyant stability of these configurations, the study of which will demand vertically-stratified global disk models.
As emphasised by \citet{2016ApJ...819...68X}, and demonstrated by \citet{2015ApJ...801...84G} in the context of AD, the field-strength dependence of diffusion coefficients, that is $\eta_{\rm H} \sim B$ and $\eta_{\rm A}\sim B^2$, opens the possibility of highly non-trivial evolution of the system once it reaches non-linear field amplitudes.
Finally, a spatial gradient in the coefficients \citep[for instance at the FUV interface as shown by][]{2015ApJ...801...84G} can produce non-negligible electromotive forces (potentially leading to significant dynamical effects) all by itself.

\subsection{Photoevaporative mass-loss}
\label{sec:photoevap}
In addition to magnetically driven winds and jets, protoplanetary disks can also lose mass through photoevaporation. Stellar UV and/or X-ray irradiation heats the disk surface to $\sim$ $10^3$--$10^4$K, and beyond some critical radius (typically 1--10AU) the hot gas is unbound and escapes as a wind \citep[e.g.][]{alexander14,gorti16}.
 Photoevaporative disk winds have now been observed directly in a number of objects \citep[e.g.,][]{ps09,sacco12}, and are probably responsible for clearing the outer disk. If a magnetic field is present, the thermal pressure can complement the magnetic acceleration to enhance mass loss and remove angular momentum (\S \ref{sec:ad}).

The impact of photoevaporative mass-loss on volatiles, and water in particular, is less clear. The wind is typically launched from the upper layers of the disk, and can entrain particles at least as large as $\mu m$-size grains \citep{takeuchi05}. Moreover, the mass-loss is concentrated at a radius ($\sim$1--10AU) comparable to the snow line, and could therefore carry away water in both gaseous and solid phases. The impact on the disk's chemical composition therefore depends critically on its radial and (especially) vertical structure, but there has been relatively little exploration of this issue in the literature. Most relevant is the work of \citet{gh06}, who showed that photoevaporation can play a role in enriching the solar system gas giants in noble gases. However, in light of our improved understanding of both mass-loss and turbulence, more detailed modelling is required if we are to assess the importance of this process more critically.

\begin{figure}[ht]
\centering
\includegraphics[width=0.8\textwidth]{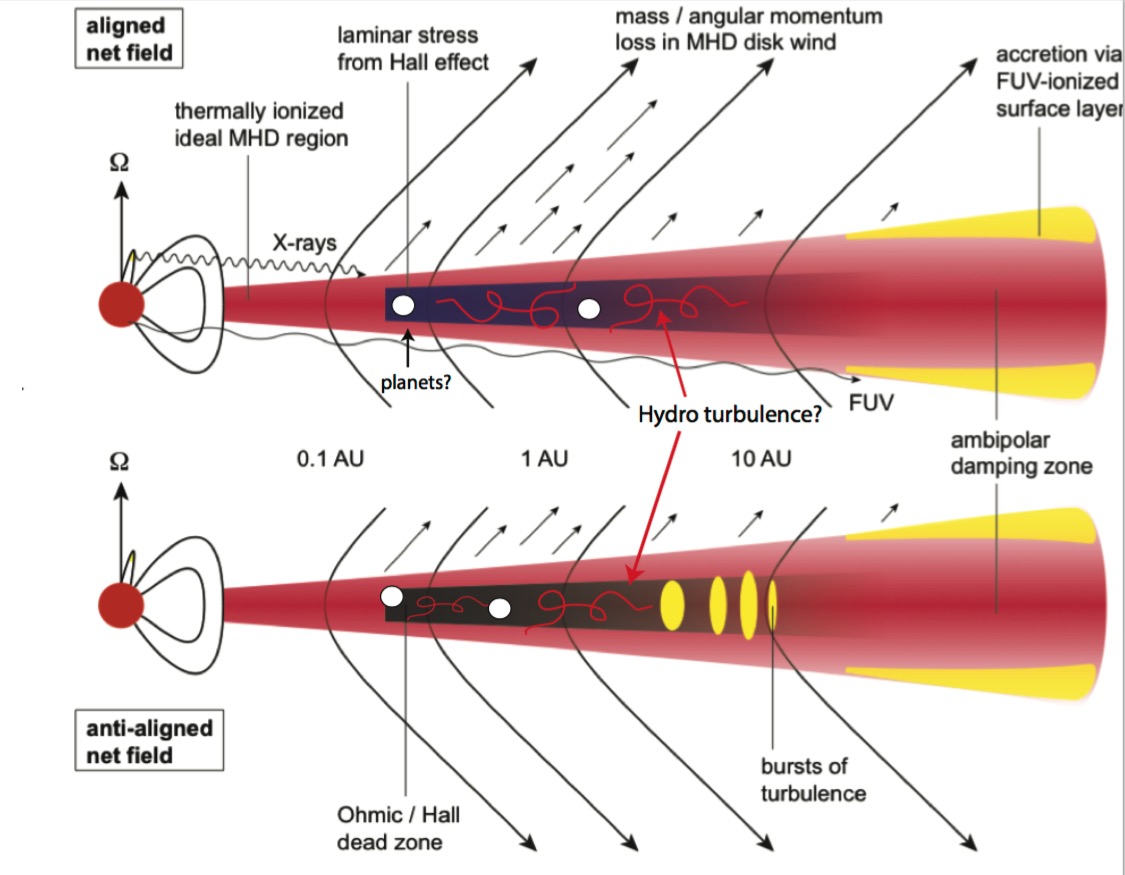}
\caption{Schematic disk structure suggested for protoplanetary disks
with non-ideal MHD transport, including the Hall effect. Modified from
Simon \etal (2015)}.
\label{fig:simonhall}
\end{figure}

\subsection{Implications for disk structure, turbulence and heating}

Figure \ref{fig:simonhall} summarizes the recent ideas about PPD structure, turbulence, and transport.  In the inner disk, thermal ionization results in MRI transport.  Beyond about 0.3 AU (for a low-mass central star), ionization levels are high only in surface layers due to the limited penetration of X-rays (and maybe cosmic rays).  Magnetic fields threading the disk can produce outflows (probably enhanced by photoevaporative heating), resulting in angular momentum transport and accretion in thin layers where AD begins to become important.  The Hall effect can produce strong azimuthal fields, and may result in qualitatively different behavior depending upon the orientation of the magnetic field relative to the angular momentum vector.

It must be emphasized that
the operation of all these non-ideal MHD processes depend upon the
amount to which the disk is able to keep magnetic flux, which is poorly constrained \citep{2014ApJ...785..127O,2014MNRAS.441..852G}, and will likely depend on the level of disk turbulence, as well as the effective magnetic Prandtl number \citep{2009A&A...507...19F} of the turbulent flow. This has helped stimulate interest in hydrodynamic
instabilities which can operate in the absence of magnetic stresses, such as the vertical shear instability \citep{2013MNRAS.435.2610N,stoll16}, the convective overstability instability \citep{2014ApJ...789...77L,2014ApJ...788...21K}, the "zombie vortex" instability (\cite{marcus15}; but see \cite{lesur16}).
These instabilities (which depend critically on rates of radiative relaxation; \cite{malygin17}) generally result in low viscosities and/or transport, with $\alpha \sim 10^{-4}$, but this could be enough to produce important effects on mass transport and dust coagulation.  Finally, it has even been suggested that, in very low viscosity disks, planets (super-Earths) can actually produce strong enough spiral waves which
reduce their inward migration while 
producing observable accretion rates
\citep{fung17}.  Obviously our
understanding of PPDs is highly uncertain; tentatively, one may conclude that both theory and observations are driving us to a picture in which most
of the disks are subject to
low levels of turbulence.

% ------------------------------------------------------------

\section{A Framework for Water Evolution in Protoplanetary Disks}

\label{sec:waterframework}

As discussed above, there are still significant uncertainties in the
extent to which the different processes described above impact how protoplanetary disks are structured and evolve.   A greater complication arises as we work to understand how all of these processes operate {\em simultaneously} to shape the distribution of water in the disk.  That is, as solids grow to different sizes, with water being incorporated into bodies with differing dynamical properties, the distribution of water ice and vapor will continuously evolve over the lifetime of the protoplanetary disk.  Here we outline a framework for understanding how this evolution may proceed, and the impact it may have on different aspects of planet formation.

Early on in disk history, dust grains will begin to coagulate and form larger and larger solids.  Outside the snow line ($r>R_{\rm SL}$), water ice-bearing solids will concentrate at the disk midplane as a result of settling, resulting in enhanced H$_{2}$O/H$_{2}$values.  Water vapor at high altitudes can diffuse downward and freeze-out on these solids, only returning again as the fine-dust that remains strongly coupled to the gas is lofted to high altitudes allowing desorption to occur again.  As the abundance of fine grained dust becomes depleted, the amount of vapor present at high altitudes in the disk diminishes, effectively lowering the height of the snow line.  This is a result of the water vapor concentration decreasing, meaning the desorption flux will exceed the adsorption flux through more of the disk \citep{krijt16b}. 

As radial drift becomes important, with solids millimeters and centimeters in size marching inwards, the location of the snow line will become blurred.  The most rapidly drifting solids, those millimeters to meters in scale, have drift timescales that are short compared to their sublimation timescale \citep{cyr98,piso15}.  As a result, icy solids may be found inside of $R_{\rm SL}$.  The exact distance to which icy solids would survive drifting inside the snow line depends on the details of the temperature gradient in the disk, but may be as much as 1-2 AU.

As these icy solids sublimate, the water will return to the gas phase.  Because the inward drift of solids can be rapid (inward drift rates will be of order cm/s to m/s relative to the gas), the delivery of water vapor to the inner disk can be much faster than the rate at which it is redistributed by radial diffusion.  This imbalance will lead to water vapor becoming enhanced in the inner disk, with concentrations increasing either until the outward flux of water vapor via diffusion equals the inward flux from drifting pebbles, or until inward drift ceases either because the pebbles are accreted by growing planets or are no longer able to form due to the depleted reservoir of solids in the outer disk \citep{cuzzizahnle04,cieslacuzzi06,estradacuzzi16}. Detailed models suggest that water vapor enhancements of a 3-5$\times$ over the ``canonical'' value could be seen inside the snow line \citep{cieslacuzzi06,estradacuzzi16} \footnote{\citet{cieslacuzzi06} used an incorrect form of the equation describing the diffusive redistribution of materials; correction of this term leads to changes in the predicted concentrations of a factor of a few.  \citet{estradacuzzi16} provided a more rigorous treatment of growth and transport and found that the qualitative relationships and behaviors hold true.}.

Once the inward drift of icy solids slows, vapor diffusion will remove water vapor by carrying it beyond the snow line to freeze-out again.  If the ice is locked up efficiently beyond the snow line so that it cannot drift inward again, then the inner disk will become continuously depleted in water vapor throughout the rest of its lifetime \citep{stevensonlunine88,cyr98,cuzzizahnle04,cieslacuzzi06}.  This diffusion serves to increase the ice concentration just beyond the snow line, meaning that the the region just beyond the snow line is expected to become enhanced in solids.  The migration of the snow
line over time, as a result of the evolution of the disk, would set exactly where such enhancements would occur.  The particulars of
disk evolution may change things, as if the net flows of gas around the snow line, depending on their orientation, could carry the water vapor enhanced/depleted gas inside the snow line outward/inward, or the ice enhanced/depleted gas inward, which would impact the overall distribution of water in the disk over time.

\subsection{Implications of Water Evolution for Planet Formation}
\label{sec:waterevolution}

The varying dynamic properties of water-bearing species will lead to continuous evolution in the distribution of water within the disk, both in the gas and solid phases.  The magnitude and timescale for this evolution will depend on many uncertain parameters, including the efficiency of particle growth and the level of turbulence (diffusion) within the disk.  There is, however, a number of features in full-born and proto-planetary systems that may be linked to this evolution which we discuss here.  By understanding these relationships, perhaps we can begin to evaluate the uncertainties in the parameters which describe protoplanetary disk processes.

The concentration of solids that would develop just outside of the snow line due to the inward drift of ices from greater heliocentric distances and the outward diffusion of ices may serve to seed the early stages of giant planet formation.  \citet{stevensonlunine88} originally suggested that the outward diffusion of vapor from the inner Solar System to beyond the assumed snow line location of 5 AU could have been a catalyst for the formation of Jupiter, as a massive core could grow more readily from the higher concentration of mass then expected.  While disk temperatures are predicted to be such for the snow line around a solar mass star to have been located closer to 1-3 AU, it is also possible that Jupiter did not form at its current orbit, and may have migrated outwards to its current location at some later point \citep[e.g.][]{walsh11}.  Further, a general view for many large exoplanets close to their host stars is that they formed further out in their protoplanetary disk then migrated inwards to where we observe them today; perhaps this was due to a very distant snow line early in disk evolution when mass accretion rates (viscous dissipation) were high (assuming there is significant viscosity; \S \ref{sec:angmom}.)

The fluctuating water conditions in the inner protoplanetary disk may play a role in shaping the chemistry and mineralogy of planetary building blocks that form there.  \citet{cuzzizahnle04} originally suggested that water vapor levels could be enhanced by many orders of magnitude by the inward drift of solids, which may have explained how iron-bearing silicates were able to form (in the absence of extreme oxygen enhancements, Fe is expected only to exist in metallic form \citep[e.g.][]{fedkin06}).  However, it appears that such enhancements are limited to just a factor of a few, due to finite reservoirs of ices and the difficulty in growing large populations of very rapidly drifting solids \citep{cieslacuzzi06,estradacuzzi16}.  Later water vapor depletions would aid in the development of very reducing conditions, and may be required to explain the mineralogy observed in enstatite chondrites \citep{cyr99,hutsonruzicka00}.

It is worth noting that the evolution described here would not be limited to water alone.  CO will freeze-out onto solids in protoplanetary disks at temperatures below $\sim$20 K \citep{qi13}, and will be subjected to the same settle/drift-adsorb/desorb cycling described above.  This will also lead to fluctuating distributions of CO in protoplanetary disks, something that must be remembered as this molecule is readily observed and used to trace gas evolution and mass \citep{williams14}.

\section{Final Remarks}

\label{sec:final}

Theoretical expectations of the positions of water snow lines
in protoplanetary disks are unclear, due to the large
uncertainties in dust opacities and the amount of viscous
dissipation present.  Unfortunately, observations so far
do not provide meaningful constraints because of limited
angular resolution and the significant dust optical
depths in inner disks. Further progress might result
from longer-wavelength interferometry to 
detect opacity transitions due to ice that arise
from the fact that water ice on the surface of grains makes them more sticky, and thus able to grow to larger sizes than the rocky grains inside of the snow line \citep{banzatti15}.  Additionally, more 
sophisticated disk models with advanced
chemistry and radiative transfer is needed
to take advantage of the current improvements in spatially-resolved dust and
molecular emission in disks to provide
further insight into the distribution of water.
For now, all that can be said with reasonable confidence is
that snow lines in protoplanetary disks around solar-type stars should lie in the range of a few to one AU, and
move in with time as accretion slows and dust opacities
decrease.

The existence of transitional disks, with large holes
cleared of small dust at early ages, suggests that
some planetesimal, embryo, or planet formation stages
might occur during optically-thin phases of inner disks.
If the formation of Jupiter could have produced such
a transitional disk structure by preventing small dust
from entering the inner Solar System, the effectively optically-thin temperature
estimate of equation (\ref{eq:tmmsn}) might be relevant
after all.

%
% For two-column wide figures use
%\begin{figure*}
% Use the relevant command to insert your figure file.
% For example, with the graphicx package use
 % \includegraphics[width=0.75\textwidth]{example.eps}
% figure caption is below the figure
%caption{Please write your figure caption here}
%\label{fig:2}       % Give a unique label
%\end{figure*}
%

\begin{acknowledgements}
The research of LH was supported by the University of Michigan and in part by NASA grant NNX16AB46G.
FC acknowledges support from NASA's Exobiology and Outer Planets Research Programs (NNX12AD59G and NNX14AQ17G).  OG has received funding from the European Research Council (ERC) under the European Union's Horizon 2020 research and innovation programme (grant agreement No 638596).
RA has received funding from the European Research Council (ERC) under the European Union's Horizon 2020 research and innovation programme (grant agreement No 681601), and also acknowledges support from the Leverhulme Trust through a Philip Leverhulme Prize.
\end{acknowledgements}

% BibTeX users please use one of
\bibliographystyle{aps-nameyear}      % American Physical Society (APS) style, author-year citations
\bibliography{refs1pdf}                % name your BibTeX data base

%
%\nocite{*}   <--- %%%%% OG: I'm removing this since otherwise un-cited references will appear in the bibliography
%

% Non-BibTeX users please use
%\begin{thebibliography}{}
%
% and use \bibitem to create references. Consult the Instructions
% for authors for reference list style.
%
% Format for Journal Reference
%\bibitem[\protect\citeauthoryear{Aamport}{1986}]{RefJ}
%L.A. Aamport, \mbox{G-Animal's} Journal \textbf{41} (7), 73 (1986).
%This is a full ARTICLE entry
%
%% Format for books
%\bibitem[\protect\citeauthoryear{Knuth}{1981}]{book-full}
%D.E. Knuth, \textit{Seminumerical algorithms}, 2nd edn.
%The Art of Computer Programming,
%vol. 2 (Addison-Wesley, Reading, 1981).
%This is a full BOOK entry
%
%
%% Format for proceedings
%\bibitem[\protect\citeauthoryear{Oz and Yannakakis}{1983}]{RefB}
%W.V. Oz, M. Yannakakis (eds.),
%in \textit{All ACM Conferences} (Academic Press, Boston, 1983).
%This is a full PROCEEDINGS entry
%% Other formats available: INPROCEEDINGS, PHDTHESIS, TECHREPORT, 
%% UNPUBLISHED, MISC, MASTERSTHESIS, MANUAL, INCOLLECTION, BOOKLET
%% etc
%\end{thebibliography}
%
\end{document}